# Helium ordered trapping in arsenolite under compression: Synthesis of $He_2As_4O_6$

J. A. Sans,[1,*] F. J. Manjón,[1] C. Popescu,[2] V. P. Cuenca-Gotor,[1] O. Gomis,[3] A. Muñoz,[4] P. Rodríguez-Hernández,[4] J. Pellicer-Porres,[5] A. L. J. Pereira,[1] D. Santamaría-Pérez,[6] and A. Segura[5]

[1] Instituto de Diseño para la Fabricación y Producción Automatizada, MALTA Consolider Team, Universitat Politècnica de València, 46022 València, Spain

[2] ALBA-CELLS, 08290 Cerdanyola, Barcelona, Spain

[3] Centro de Tecnologías Físicas, MALTA Consolider Team, Universitat Politècnica de València, 46022 València, Spain

[4] Departamento de Física, Instituto de Materiales y Nanotecnología, MALTA Consolider Team, Universidad de La Laguna, 38205 La Laguna, Tenerife, Spain

[5] ICMUV-Departamento de Física Aplicada, MALTA Consolider Team, Universitat de València, 46100 Burjassot, Spain

[6] Earth Sciences Department, University College London, Gower Street, WC1E 6BT, London, UK

**Abstract**

The compression of arsenolite (cubic $As_2O_3$) has been studied from a joint experimental and theoretical point of view. Experimental X-ray diffraction and Raman scattering measurements of this molecular solid at high pressures with different pressure-transmitting media have been interpreted with the help of *ab initio* calculations. Our results confirm arsenolite as one of the softest minerals in absence of hydrogen bonding and provide evidence for helium trapping above 3 GPa between adamantane-type $As_4O_6$ cages, thus leading to a new compound with stoichiometry $He_2As_4O_6$. Helium trapping alters all properties of arsenolite. In particular, pressure-induced amorphization, which occurs in pure arsenolite above 15 GPa, is impeded when He is trapped between the $As_4O_6$ cages; thus resulting in a mechanical stability of $He_2As_4O_6$ beyond 30 GPa. Our work paves the way for the modification of the properties of other molecular solids by compression depending on their ability to trap relatively small atomic or molecular species and form new compounds. Furthermore, our work suggests that compression of molecular solids with noble gases as helium could result in unexpected results compared to other pressure-transmitting media.

Molecular solids are very soft materials with open framework structures composed of molecular units, exhibiting strong covalent interatomic bonds, which are linked by weak forces of van der Waals or hydrogen type (intermolecular bonds). Pressure is a thermodynamic variable which allows tuning interatomic distances and consequently is a powerful tool to study the atomic interactions and the connectivity of different molecular units of molecular crystals. This fact lead to the study of the vibrational properties of many molecular solids in the 1970's; i.e., the early days of Raman scattering (RS) measurements at high pressures (HP) using a diamond anvil cell (DAC).[1-5] The aim of those studies was to easily compact molecular solids with relatively small pressures which could be able to produce strong changes in molecular interactions and induce the occurrence of phase transitions. In the 1980's, many HP-RS studies of molecular solids at relatively small pressures were already reviewed.[6] In those early days of HP-RS studies, HP X-ray diffraction (XRD) measurements using high flux sources, like synchrotrons, could not be performed to obtain the most accurate information about the effect of pressure on the interatomic distances of materials. In fact, X-ray diffractometers adapted to use DACs were only available in a few laboratories all over the world. Furthermore, most of those HP experiments were performed either without a pressure-transmitting medium (PTM) or using a liquid fluid (Octoil-S, water, or (4:1) methanol-ethanol mixture) as a PTM at pressures not exceeding 10 GPa. Finally, in those early days of HP-RS studies, *ab initio* calculations were not available for the study of molecular crystals, which are complex materials difficult to simulate because they feature both strong intramolecular and weak intermolecular bonds. Therefore, it was difficult to interpret the changes of materials properties caused by the strong increase of interatomic interactions observed in compressed molecular crystals. In this scenario, many interesting results of the effect of pressure on molecular solids came up in the 1970's and early 1980's which unfortunately resulted in rather incomplete studies of the effect of pressure and the different PTM on the structure and properties of molecular solids.

Arsenic oxide ($As_2O_3$) belongs to the sesquioxide family of group-15 elements, which also includes $P_2O_3$, $Sb_2O_3$ and $Bi_2O_3$. In particular, $As_2O_3$ crystallizes either in a cubic structure [space group (SG) 227, Fd-3m, Z=16] (named arsenolite)[7] or in monoclinic structures [SG 14, $P2_1/c$, Z=4] (named claudetite I and claudetite II)[8-10]; however, it can also be obtained in an amorphous (glass) phase.[11,12] A number of polymorphs are also known for $Sb_2O_3$ and $Bi_2O_3$ at ambient conditions.[13] Most of the polymorphs of group-15 sesquioxides show close structural connections since many structures could be derived from a defective fluorite structure through symmetry breaking and local distortions.[14,15] Some of those polymorphs constitute molecular solids, like arsenolite ($\alpha$-$As_2O_3$) and senarmontite ($\alpha$-$Sb_2O_3$), which are the most stable phases of $As_2O_3$ and $Sb_2O_3$. Arsenolite is known to be a very soft mineral with an intermediate hardness between that of Talc ($Mg_3Si_4O_{11} \cdot H_2O$) and Gypsum ($CaSO_4 \cdot 2H_2O$); being these two minerals the softest materials in the Moss scale. Arsenolite, as the other known polymorphs of arsenic oxide, is composed of pseudo–tetrahedral units consisting of an arsenic atom surrounded by three oxygen ligands and a lone electron pair (LEP). Unlike the other arsenic oxide polymorphs, pseudo-tetrahedra in arsenolite are configured in closed-compact adamantane-type $As_4O_6$ molecular cages bonded together due to weak van der Waals forces (see **Fig. 1a**); thus forming a *true* molecular solid. It must be stressed that among sesquioxides and sesquichalcogenides of group-15 elements, the formation of molecular cages is also found in cubic $Sb_2O_3$ (or $Sb_4O_6$) at room temperature and cubic $P_2O_3$ (or $P_4O_6$) at low temperatures, as well as in $S_4N_4$, $P_4S_4$, $As_4S_4$, $As_4S_3$ and $As_4S_5$. The reason for the molecular character of these materials is the strong activity of LEPs in group-15 elements and also in O and S.[16]

$As_2O_3$ is used as a raw material for the production of other inorganic arsenic compounds, alloys, and organic arsenic compounds. In particular, arsenolite is a transparent molecular solid (white powder) which is highly toxic and must be handled with care. In the antiquity, this compound became famous because it was employed in several murders as homicidal agent. However, nowadays $As_2O_3$ has attracted great attention since it has been introduced in small proportions in chemotherapy in order to improve the success rate of cured patients in the treatment of acute promyelocytic leukaemia.[17-18] Therefore, the understanding of the physical-chemical properties of the different polymorphs observed in sesquioxides of group-15 elements is very important for many technological applications and high-pressure studies are crucial to clarify the relationships between the different polymorphs and their different properties. In this respect, HP-RS and HP-Fourier Transform Infrared (FTIR) measurements of arsenolite[19] (using CsI as PTM), and HP-RS measurements of arsenolite, claudetite, and glass[20,21] (without PTM) have been performed. On the other hand, HP-XRD measurements of arsenolite, claudetite and glass $As_2O_3$ have been reported[20-23]; however, the pressure dependence of the atomic structure of arsenolite and its equation of state (EOS) have not been properly addressed in previous works and the effect of different PTM on the properties of these polymorphs under compression has not yet been studied.

In this work we provide a comprehensive understanding of the compression of arsenolite using different PTM (silicone oil, 4:1 methanol-ethanol mixture and He) up to pressures of 30 GPa and without PTM up to 12 GPa. Our study combines two experimental techniques (XRD and RS) and *state-of-the-art ab initio* calculations providing structural, elastic, electronic and lattice dynamical properties of arsenolite at different pressures. Our experiments show a different behavior of arsenolite under compression when He is used as a PTM. Our calculations are able to explain those results of arsenolite compressed with He only if He atoms are located at a defined Wyckoff site (16d) inside arsenolite; i.e., as if a new compound with $He_2As_4O_6$ stoichiometry is formed due to He trapping between the molecular $As_4O_6$ cages of arsenolite. We will show that the formation of the new compound at HP due to He trapping alters the structural, vibrational, electronic and elastic properties of arsenolite. These results suggest that the modification of the properties of other molecular solids (without a change of structure) can be explored when compressed with small elements or molecules able to enter into the framework of the molecular solid.

Experimental XRD patterns are provided in the supplementary material (**Fig. S1**). Those measurements reveal a progressive increase of the angle for all diffraction peaks of the cubic structure with pressure, as expected for a decrease in the unit cell volume under compression. The absence of new peaks at high pressures clearly indicates that no phase transition occurs along the pressure range studied. In the case of the experiment with silicone oil, the increase of the intensity of the (200) Bragg peak and a decrease of the relative intensity of the peak corresponding to the lowest angle (111) reveal an increase of preferential orientation along the c-axis with increasing pressure. This preferential orientation has been considered during the Rietveld refinement by the addition of the spherical harmonics preferential orientation coefficients giving a texture index around 1.8-2, which is considered an intermediate value between 1 (no texturized) and 3 (strongly texturized). This feature is not observed in the rest of the experiments using He, (4:1) methanol-ethanol as PTM or no PTM.

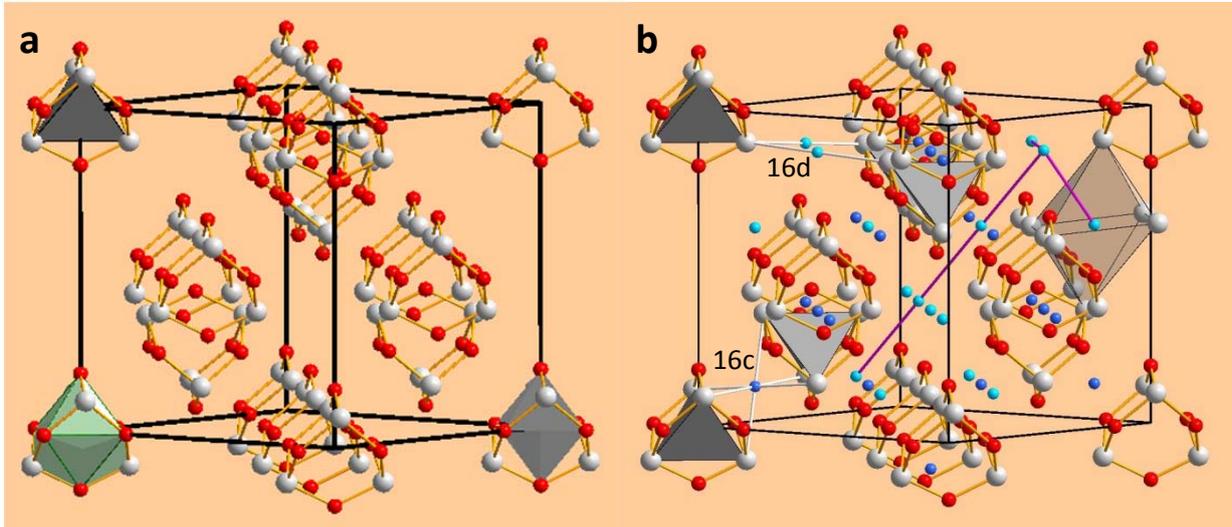

**Figure 1 | Crystalline structure of pure arsenolite and arsenolite with Helium.** Detail of the cubic unit cell of arsenolite ($As_4O_6$) at low pressures: **a**, pure arsenolite, and **b**, He-trapped arsenolite with He inserted into the two possible Wyckoff sites (16c and 16d). Big gray balls, medium-size red balls and small dark and light blue balls represent As, O, He(16c) and He(16d) atoms, respectively. Pink lines correspond to 3D diffusion paths for He along 16d sites.

A notable broadening of Bragg peaks above 15 and 20 GPa were observed in the experiments without PTM and with silicone oil or methanol-ethanol, respectively; which can be ascribed to the onset of pressure-induced amorphization (PIA). This was corroborated by comparing the XRD patterns of samples before and after compression which reveal that recovered samples does not maintain the same crystalline quality after the pressure cycle. This result is in good agreement with previous results obtained in arsenolite compressed without any PTM.[21] Noteworthy, arsenolite compressed with He shows no evidence of PIA up to the maximum pressure reached (29.4 GPa), even though there is a small increase of the width of the Bragg peaks at the highest pressures, likely due to a partial loss of hydrostaticity. In this case, the released sample after the full pressure cycle shows the cubic arsenolite phase with crystalline quality comparable with the sample at ambient conditions, which is a clear signature of the absence of PIA (see **Fig. S1**).

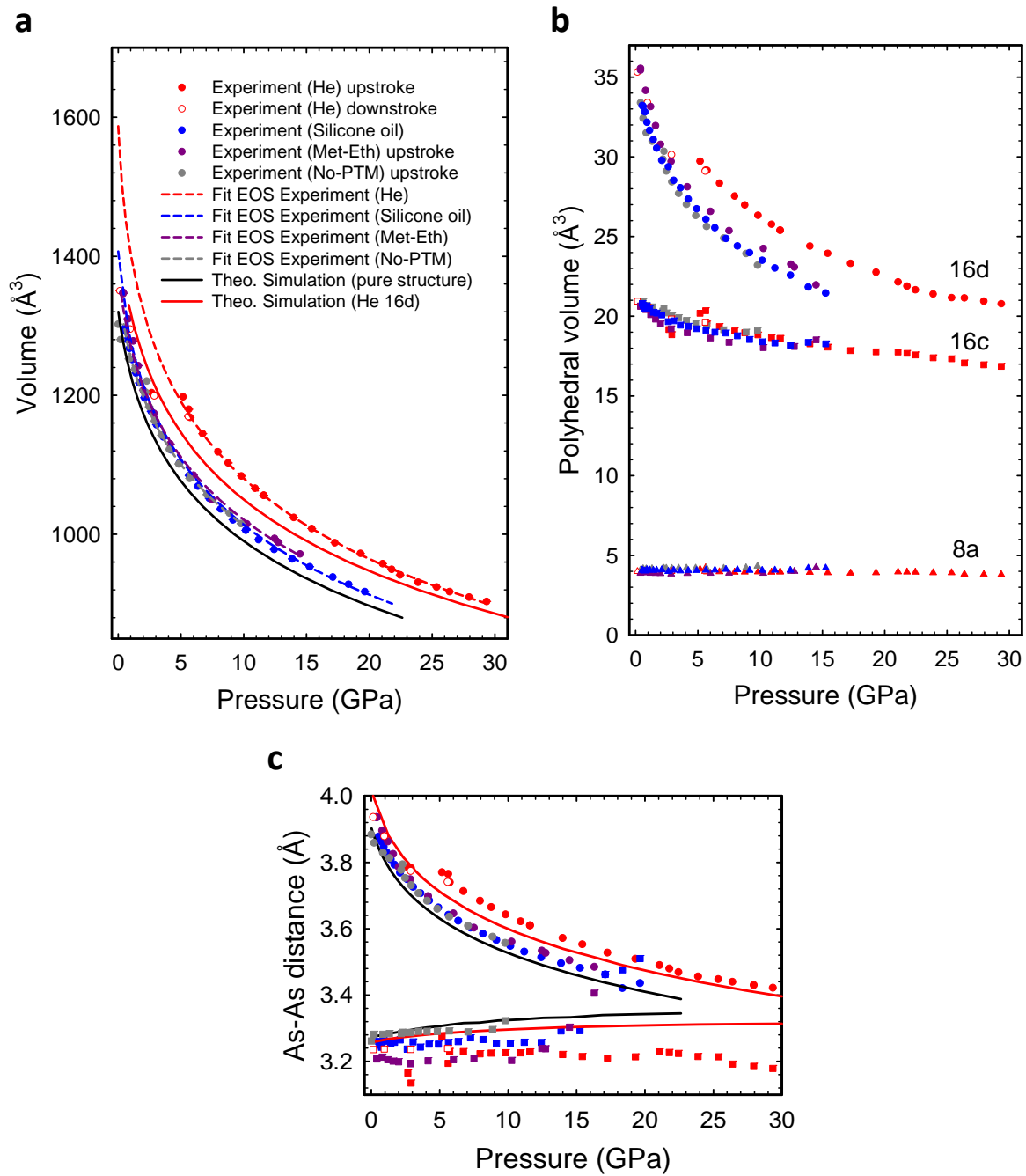

**Figure 2 | Evolution of the structural parameters of arsenolite under pressure with different PTM.** Grey, blue, purple, and red symbols correspond to data of arsenolite compressed without PTM, with silicone oil, with methanol-ethanol mixture and with He, respectively. Black (red) solid line corresponds to theoretical data of pure (He-inserted in 16d sites) arsenolite. Dashed lines correspond to fit of experimental data. **a**, equation of state of arsenolite compressed with different PTM. **b**, compression of the polyhedral volume around different Wyckoff crystallographic sites: 16d (circles), 16c (squares) and 8a (triangles). **c**, compression of the shortest arsenic-arsenic interatomic distances belonging to the same molecular unit (squares) and different molecular units (circles).

The pressure dependence of the atomic parameters of arsenolite (**Fig. S2** of the supplementary material) was obtained by Rietveld refinement along a large pressure range due to the relatively high quality of our experimental XRD patterns.[24] The compression of arsenolite's volume with increasing pressure is plotted in **Fig. 2a**. The experimental EOS obtained in the experiments using no PTM and silicone oil or methanol-ethanol as PTM yield a bulk modulus around 7(2) GPa which is in good agreement with the theoretical EOS for pure arsenolite.[25] A similar bulk modulus can be inferred from data of arsenolite compressed with He below 4 GPa.[26] Arsenolite's bulk modulus is much smaller than those of other sesquioxides (see **Table SI** in supplementary material). In general, the bulk modulus of sesquioxides of group-15 elements ($As_2O_3$, $Sb_2O_3$, $Bi_2O_3$) is well below 80 GPa,[27,28] whereas sesquioxides of group-13 elements in the same row of the periodic table ($Ga_2O_3$, $In_2O_3$, $Tl_2O_3$) is well above 160 GPa.[29-31] The reason for the small bulk moduli of sesquioxides of group-15 elements is the presence of the cation LEP which favors the formation of voids in the structural units resulting in open-framework low-compact structures.

It must be stressed that arsenolite's small bulk modulus is in the range of the softest molecular crystals.[23,32-35] Arsenolite's molecular units bear resemblance with those of mineral realgar ($\alpha$-$As_4S_4$) and pararealgar ($\beta$-$As_4S_4$).[36] They all feature a cage formed by a tetrahedron of As atoms surrounded by anions in a closed-compact configuration. Arsenolite's bulk modulus is smaller than those of $\alpha$-$As_4S_4$ ($B_0$= 8.1 GPa and $B_0'$=9.0)[34] and $\beta$-$As_4S_4$ ($B_0$= 10.9 GPa and $B_0'$=8.9).[34] In all these compounds, the geometry of the cage-like structure is based on As tetrahedral units (see **Fig. 1a**) which contributes to their high structural stability. Note that LEPs of As are oriented toward outside the cages so the large compression of the unit cell at low pressures is mainly governed by the compression of the empty space between the cages linked by weak van der Waals bonds.

The pressure dependence of the unit cell volume of arsenolite shows a striking behavior above 3 GPa when compressed with He (see **Fig. 2a**). Volume decreases in a normal fashion below 3 GPa; however, the volume at 5 GPa is almost the same as at 3 GPa. Then, above 5 GPa arsenolite's unit cell volume decreases with increasing pressure in a monotonous way. A fit of the EOS of arsenolite compressed with He with data above 5 GPa yields a bulk modulus of 4 GPa (see **Table I**). These results can be interpreted as a signature of He entry in arsenolite's structure between 3 and 5 GPa and the consequent He trapping between $As_4O_6$ cages above 5 GPa. He entry and trapping in arsenolite would explain the similar volume observed around 3 and 5 GPa due to compensation of the decrease of volume with pressure and the increase of volume with He entry. We will show that the small variation of compressibility of arsenolite after He entry is due to the fact that He is located at the largest voids (16d Wyckoff sites) of the structure.

Arsenolite's compressibility can be described in terms of the compressibility of their constituting polyhedral units: i) adamantane-type $As_4O_6$ cage, centered around the 8a (0,0,0) Wyckoff site, ii) the quasi-octahedral unit centered around the 16c site and iii) the octahedron centered around the 16d site (see **Fig. 1b**). He trapping can occur in all these polyhedral units but preferentially in 16c and 16d sites. The similar compressibility of pure arsenolite and He-trapped arsenolite above 5 GPa suggests that He trapping does not alter significatively the interatomic bonds. In order to understand how He affects arsenolite's compressibility, the experimental compressibility of the different polyhedra with different PTM are plotted in **Fig. 2b**.[37] It can be observed that there is no obvious change in the compressibility of the smaller polyhedral units around 8a and 16c crystallographic sites with the different PTM used. This result suggests that He does not enter in these small polyhedral units and particularly in 16c sites, which feature the smallest distances between neighbor adamantane-cages. Note that the closest intermolecular As-O distance is below 3 Å in the whole pressure range till 30 GPa (see **Fig. S3b**), but well above the van der Waals radius of He (1.4 Å).[38] The most striking evidence for He trapping in arsenolite at high pressure is given by the different pressure dependence of the volume of the octahedron around the 16d Wyckoff site depending on the PTM used. The jump in volume of this polyhedral unit between 3 and 5 GPa is a clear evidence of He entry in 16d sites and explains the jump in the unit cell volume observed in this pressure range.

**Table I | Equation of state of arsenolite**

|  | $V_0$ (Å$^3$) | $B_0$ (GPa) | $B_0'$ |
|---|---|---|---|
| **Experimental (He)** | 1587(26) | 4(2) | 12.9 |
| **Experimental (Sil. Oil)** | 1407(2) | 6(1) | 12.9 |
| **Experimental (Met-Eth)** | 1385(15) | 7(2) | 12.9 |
| **Experimental (No-PTM)** | 1361(20) | 7(2) | 12.9 |
| Theo. Sim. (He 16d) | 1467(1) | 4.8(3) | 15.6(7) |
| Theo. sim. (pure) | 1331(2) | 7.6(3) | 12.9(3) |

Furthermore, in order to verify the hypothesis of He entry and trapping in arsenolite above 3 GPa, we have theoretically simulated the compression of arsenolite with He located in the 8a, 16c and 16d Wyckoff positions or in several of them simultaneously (see **Fig. S2** of the supplementary material). Despite the interpretation of bulk compressibility data in order to understand where He is trapped is not as straightforward as that of polyhedral volume compressibility, after a careful analysis we have concluded that our experimental results can be better explained by considering that He is mainly trapped in the 16d sites of arsenolite (**Fig. 1b**) in good agreement with the compressibility of the polyhedral units shown above. Note that a rather good agreement between the theoretical and experimental behavior of arsenolite compressed with He in the range between 5 and 30 GPa is found (see **Fig. 2a**). Therefore, our joint experimental and theoretical study clearly point to the entry and trapping of He in 16d sites of arsenolite above 3 GPa.

Additional proofs which confirm He trapping in arsenolite are provided by HP-RS measurements using the similar PTM as those used in XRD measurements. Experimental RS spectra using no PTM and (4:1) methanol-ethanol mixture or helium as PTM are provided in the supplementary material (**Fig. S4**). When using no PTM or (4:1) methanol-ethanol mixture as PTM, HP-RS spectra are similar to those already reported[21,36] and the pressure dependence of the frequencies of the first-order Raman-active modes are well described by our theoretical calculations for pure arsenolite (see **Fig. 3a**); however, a different pressure dependence for the frequencies of several Raman-active modes is observed when He is used as PTM (see **Fig. 3b**). Below 3 GPa Raman mode frequencies follow the same pressure dependence as in the case without PTM or (4:1) methanol-ethanol mjixture; however, above 3 GPa almost all Raman modes suffer a small shift in frequency and some of them show a dramatically different behavior with pressure. In fact, the sign of the frequency shift for each mode is consistent with a volume increase at around 3 GPa: for modes with a positive pressure coefficient the shift is negative and viceversa.

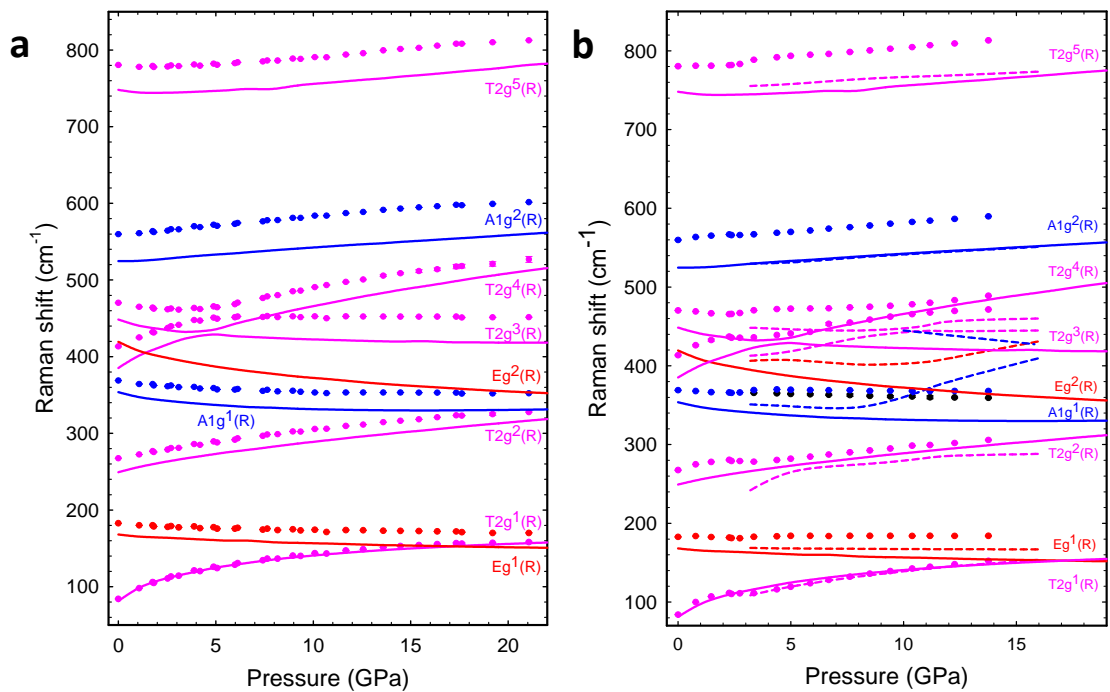

**Figure 3 | Pressure dependence of Raman-active mode frequencies of arsenolite with different PTM**. Experimental (symbols) pressure dependence of first-order Raman-active mode frequencies using: **a**, (4:1) Methanol-Ethanol as PTM and **b**, He as PTM. Solid lines correspond to calculations of pure arsenolite while dashed lines correspond to calculations of arsenolite with He inserted in 16d Wyckoff sites. A different pressure dependence of Raman modes is observed in He compressed arsenolite above 3 GPa.

The entry of He above 3 GPa is coincident with the splitting of the soft $A_{1g}^1$ mode into two modes. However, the most striking case of different behavior is that of the two $T_{2g}$ modes between 400 and 500 cm$^{-1}$ corresponding to As-O bending modes inside $As_4O_6$ cages. These two modes undergo a phonon anticrossing around 4.5 GPa when (4:1) methanol-ethanol mixture is used as PTM. The anticrossing is very well described by our lattice dynamics calculations of pure arsenolite. Similar results can be observed in the experiment without PTM (see **Fig. S5**). On the other hand, when arsenolite is compressed with He, these two modes undergo a considerably splitting above 3 GPa and the anticrossing is retarded to 11 GPa. Notably, these experimental results are theoretically reproduced when He is inserted in 16d Wyckoff sites (see **Fig. 3b**). In summary, our experimental HP-XRD and HP-RS results and our theoretical results confirm that He incorporates above 3 GPa into arsenolite at ordered positions leading to a new compound, likely of stoichiometry $He_2As_4O_6$, whose properties are strikingly different to those of pure arsenolite.[39]

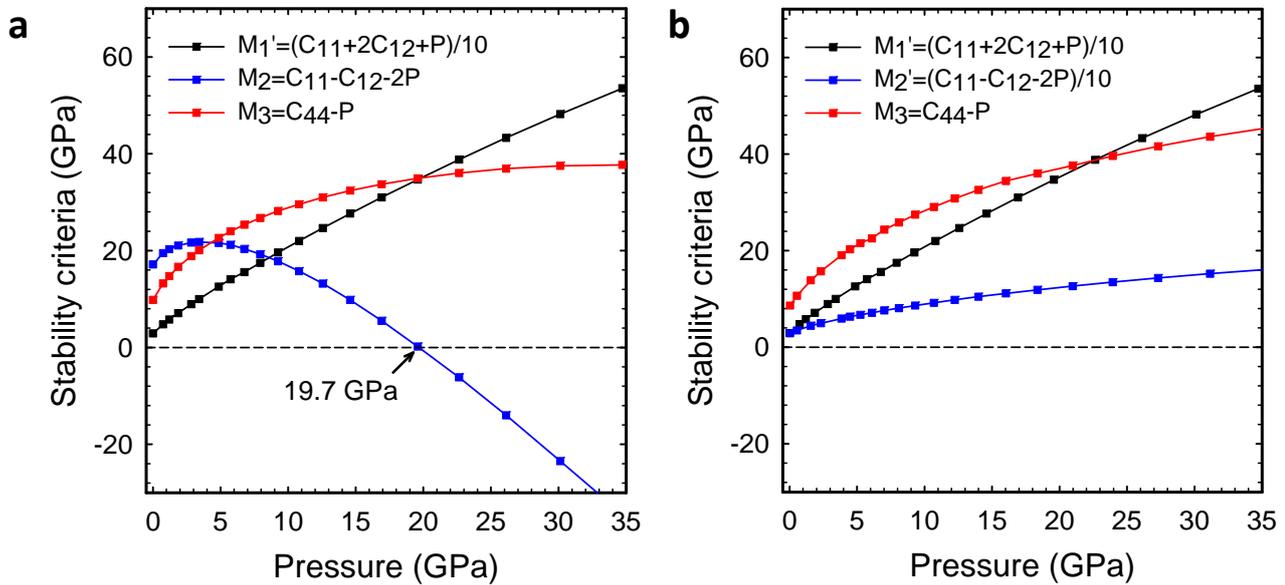

**Figure 4 | Mechanical stability criteria under pressure.** Generalized Born criteria for mechanical stability under hydrostatic pressure: **a**, pure arsenolite and **b**, $He_2As_4O_6$ with He in 16d sites.

Finally, we want to stress that the new compound formed above 3 GPa has different properties from pure arsenolite. A considerable change in the elastic, electronic and optical properties of $He_2As_4O_6$ is predicted compared to $As_4O_6$. Our calculations for pure arsenolite suggest a strong decrease of the indirect bandgap from 4 to 3 eV between 1 atm and 15 GPa. However, a similar indirect bandgap around 4 eV, with a very small increase as a function of pressure between 5 and 30 GPa, is obtained for $He_2As_4O_6$. As regards the elastic properties, He-trapping leads to a complete change of the structural stability of arsenolite. As already commented, HP-XRD measurements locate the onset of PIA in arsenolite above 15 GPa when compressed with PTM different from He; however, no sign of PIA is observed in He-compressed arsenolite even to 30 GPa. In order to verify whether or not He trapping in 16d sites could enlarge the structural stability pressure range of arsenolite we have conducted theoretical calculations of the pressure dependence of the elastic constants in pure arsenolite and $He_2As_4O_6$ (see **Fig. S6** in supplementary material). From the calculated elastic constants one can obtain the stiffness coefficients and apply the generalized Born criteria to studty the mechanical stability at HP of pure arsenolite and $He_2As_4O_6$ (see **Fig. 4**). For pure arsenolite, $M_2$ mechanical stability criterion is violated at 19.7 GPa (see **Fig. 4a**). This pressure is close to the pressure at which the onset of PIA process occurs according to XRD measurements when using no He as a PTM. On the other hand, $M_2$ mechanical stability criterion is not violated even at 30 GPa in $He_2As_4O_6$ as observed in **Fig. 4b**. This result is in very good agreement with our experimental results and, on one hand, supports He trapping in arsenolite above 3 GPa leading to the formation of a new compound, and on the other hand, confirms the change of the properties of He-trapped arsenolite compared to pure arsenolite.

The different stability of the cubic phase of $As_4O_6$ at HP depending on the PTM used can be understood on the basis of the evolution of the As-As distances with increasing pressure (see **Fig. 2c**). In $As_4O_6$, there are two characteristic As-As distances: i) one corresponding to the distance between As cations inside the tetrahedron of the cage-like molecular unit (internal As-As distance), and ii) one corresponding to the minimum distance between As atoms of two neighboring cages (external As-As distance). In pure arsenolite under compression, the external As-As distance tends to decrease monotonously under compression due to the closing of the gap among $As_4O_6$ molecules. On the other hand, the internal As-As distance exhibits a slight increase with pressure; i.e., the arsenic tetrahedra shown in **Fig. 1** increase in size with pressure. This surprising result is likely due to an increase of the electric charge of the $As_4O_6$ cage due to a charge transfer from As and O LEPs to the internal bonds of the $As_4O_6$ molecule. This charge transfer is favored by the strong compression of the gap among the cages governed by weak van der Waals forces. Extrapolation of both internal and external As-As distances suggests that both tend to become equal around 25 GPa. In fact, the external As-As distance is less than 3% longer than the internal As-As distance around 20 GPa. The close similarity of external and internal As-As distances above 15 GPa results in strong intermolecular interactions that unstabilize the cubic structure of arsenolite. On the other hand, He-trapping in arsenolite above 3 GPa leads to a considerably increase of the external As-As distance (see the jump in **Fig. 2c**) and consequently the compound tends to avoid the intermolecular interactions which unstabilize the crystalline structure of arsenolite and extend the stability range of the structure beyond 30 GPa. In fact, the experimental external As-As distance is still 6% larger than the internal As-As distance at 30 GPa. This difference in external As-As distances (and also in external As-O distances (see **Fig. S3**) in He-trapped arsenolite with respect to its pure arsenolite counterpart is the responsible for the different intermolecular interactions reflected in the vibrational RS spectra of pure and He-trapped arsenolite under compression. This result implies that the stability of the molecular structure of arsenolite is clearly related to the steric repulsion between the cationic As sublattice governing the formation of the cage-like structure.

Finally, the difference in the behavior depending on the PTM is assigned to the peculiar crystalline structure of arsenolite. Small He atoms can be diffused along the whole structure thanks to the existence of big connected spaces between 16d sites without the need to enter in 16c sites (see paths between He atoms at 16d sites in **Fig. 1b**). However, methanol-ethanol mixture and silicone oil cannot be inserted into the arsenolite structure because of their high molecular size compared to the diameter of the space around 16d sites. In these cases, $As_4O_6$ units cannot find any obstacle in the compression and approach to each other favoring a strong interaction among molecular units which breaks down the molecular network and lead to PIA while He-entrance and subsequent He-trapping helps to stabilize the structure keeping molecular units separated and avoiding PIA.

In summary, we have proved that arsenolite is one of the molecular solids with the smallest bulk modulus (around 7(2) GPa) in absence of hydrogen bonding and we have demonstrated He entry and trapping in $As_4O_6$ beyond 3 GPa. He trapping in arsenolite results in the formation of a new compound with stoichiometry $He_2As_4O_6$ (only stable above 3 GPa) whose properties are different from those of pure arsenolite. In particular, the new compound has a mechanical stability beyond 30 GPa while pure arsenolite undergoes PIA above 15 GPa. The present results pave the way to explore the modification of the properties of other molecular solids by incorporating small atomic or molecular species with the help of pressure. Furthermore, these results suggest that compression of molecular solids with noble gases, like helium, could result in unexpected results compared to other pressure-transmitting media due to entry of small gas molecules into the empty spaces of molecular compounds.

## Methods

**Synchrotron-based angle dispersive X-ray diffraction under pressure.** Highly pure arsenolite ($As_4O_6$) powder (99.999%) was commercially obtained from Sigma-Aldrich Company. High-pressure angle-dispersive X-Ray Diffraction (XRD) experiments at room temperature (RT) up to 30 GPa were conducted in a membrane-type diamond anvil cell (DAC) using commercial powder crushed in a mortar with a pestle to obtain a micron-sized powder. Measurements were performed with silicone oil (Rhodorsil 47V1000), methanol-ethanol (4:1 ratio) mixture, or He gas as quasihydrostatic PTM and also without any PTM. Pressure inside the DAC was estimated from the EOS of copper.[40] Experiments were performed at the BL04-MSPD beamline of the ALBA synchrotron with an incident monochromatic wavelength of 0.4246 Å focused to 20 x 20 μm$^2$.[41] Pinhole of 50 μm was used to clean the x-ray beam tail. Images covering a 2θ range up to 20° were collected using a SX165 CCD located at 240 mm from sample. One-dimensional diffraction profiles of intensity as a function of 2θ were obtained by integration of observed intensities with the Fit2D software.[42] Lattice parameters of powder XRD patterns were obtained with Rietveld refinements performed using GSAS program package.[43,44] Interatomic distances were extracted thanks to VESTA software.[45]

**Raman scattering under pressure.** Raman scattering (RS) measurements at RT excited either with 532.0 or 632.8 nm laser lines and laser power below 10 mW were performed in a backscattering geometry using a Horiba Jobin-Yvon LabRam HR UV spectrometer in combination with a thermoelectrically-cooled multichannel CCD detector (resolution below 2 cm$^{-1}$). High-pressure RS measurements in power up to 33 GPa were performed inside the DAC. Measurements were performed with the same PTM as XRD measurements except for silicon oil. A few ruby balls of about 2 μm in diameter evenly distributed in the pressure chamber were employed as a pressure sensor.[46] RS measurements were analysed by fitting Raman peaks with a Voigt profile fixing the Gaussian line width (1.6 cm$^{-1}$) to the experimental setup resolution.[47]

***Ab-initio* calculations**. *Ab initio* total-energy calculations of arsenolite were performed within the framework of density functional theory (DFT).[48] Vienna Ab initio Simulation Package (VASP) was used to carry out calculations with the pseudopotential method and the projector augmented wave (PAW) scheme, which replace the core electrons, make smoothed pseudovalence wave functions and take into account the full nodal character of the all-electron charge density in the core region.[49] Exchange and correlation term was computed through PBE for solids prescription.[50] Lattice-dynamics calculations at the zone center (Γ point) of the Brillouin zone were performed using the direct force constant approach.[51] The elastic constants can be obtained by computing the macroscopic stress for a small strain with the use of the stress theorem.[52] In the present work, we performed the evaluation of the elastic constants of arsenolite as implemented in the VASP package.[53]

**Acknowledgements**
Financial support from the Spanish Consolider Ingenio 2010 Program (Project No. CSD2007-00045) is acknowledged. The work was also supported by Spanish MICCIN under projects MAT2014-46649-C4-1/2/3-P and from Vicerrectorado de Investigación de la Universitat Politècnica de València under projects SP20140701 and SP20140871. Supercomputer time has been provided by the Red Española de Supercomputación (RES) and the MALTA cluster. J. A. S. acknowledges Juan de la Cierva fellowship program for his financial support. A. M. and P. R.-H. want to acknowledge to S. Muñoz Rodríguez for providing us with a data parsing application. A.L.J.P. acknowledges financial support through Brazilian CNPq (Project Nº 201050/2012-9)

**Author contributions**
J.A.S., C.P., O.G., F.J.M., V.P.C-G., A.L.J.P. and D.S-P. carried out the Synchrotron-based X-ray diffraction under pressure experiments. J.A.S and F.J.M. analyzed the crystal structures and diffraction patterns. V.P.C-G, J.A.S., J.P-P and F.J.M performed and analyzed the RS measurements under pressure. A.M. and P.R-H. conducted the DFT calculations; O.G. analyzed and interpreted the evolution of the elastic constants with pressure and mechanical stability criteria; J.A.S. and F.J.M. wrote the manuscript and all authors discussed the experiments the final manuscript.

**Additional information**